\documentclass[%
 reprint,
 amsmath,amssymb,
 aps,
pra,
]{revtex4-2}
\usepackage{hyperref}
\usepackage{physics}
\usepackage{subcaption}
\usepackage{ragged2e}
\usepackage{graphicx}
\usepackage{dcolumn}
\usepackage{bm}

\begin{document}

\title{Generalized fusions of photonic quantum states using static linear optics}

\author{Frank Schmidt}
 \affiliation{Institute of Physics, Johannes-Gutenberg University of Mainz, Staudingerweg 7, 55128 Mainz, Germany}
\author{Peter van Loock}
 \email[]{loock@uni-mainz.de}
\affiliation{Institute of Physics, Johannes-Gutenberg University of Mainz, Staudingerweg 7, 55128 Mainz, Germany}

\begin{abstract}
Using numerical simulations, we explore generalized fusion measurements, extending them in three key ways. We incorporate ancilla-boosting and code-boosting to optimize fusion measurements, which may, but do not necessarily, correspond to Bell measurements. For ancilla-boosting beyond Bell measurements, we identify a hierarchy of fusion efficiencies involving up to four single-photon ancillas. From this numerical evidence, we conjecture the efficiencies for a larger number of single-photon ancillas. Interestingly, our findings suggest that achieving unit efficiency is impossible without utilizing entangled ancilla states. Additionally, in the context of code-boosting, we present an example of a global measurement that surpasses the best-known, via transversal Bell measurements achievable efficiency for a small instance of the quantum parity code.
\end{abstract}

\maketitle


\section{Introduction}

With the recent surge of interest in quantum computing, there has been a parallel focus on creating large entangled states, which are essential for measurement-based \cite{mbqc} or fusion-based quantum computing~\cite{fbqc} (MBQC/FBQC). A common implementation thereof involves encoding qubits in single photons using the dual-rail encoding scheme. Unfortunately, deterministic entanglement operations on dual-rail qubits cannot be achieved using only linear optics. To introduce the necessary non-linearity, probabilistic measurements are typically employed. As a result, it becomes crucial to explore the achievable probabilities of success depending on the available ancilla resources.

More than two decades ago, Calsamiglia and Lütkenhaus~\cite{Calsamiglia2001} demonstrated that linear-optical unambiguous state discrimination of equiprobable Bell states can only succeed with a maximum probability of \(\frac{1}{2}\). Moreover, they showed that adding ancilla modes in the vacuum state does not improve this probability, and deterministic measurements are impossible with any ancilla state~\cite{PhysRevA.59.3295}, even when feed-forward operations (conditional dynamics) are included~\cite{PhysRevA.59.3295, PhysRevA.69.012302}. Only asymptotically can a 100\%-efficient Bell-state discrimination and measurement be achieved by employing a sufficiently large entangled multi-photon ancilla state, with feed-forward included~\cite{Knill2001} or excluded~\cite{Grice_BM}.
More specifically, Grice~\cite{Grice_BM} and subsequently, Ewert and van Loock~\cite{Ewert_BM} showed that the efficiency of Bell measurements can be increased arbitrarily close to unity by employing increasingly complex ancilla states. In particular, Ref.~\cite{Ewert_BM} presented a scheme achieving an efficiency of \(\frac{3}{4}\) using only four unentangled photons as the ancilla state. The optimality of these schemes has been further explored through analytical bounds and numerical simulations in Ref.~\cite{PhysRevA.98.042323}.

Given that photon loss is a significant challenge in the use of photonic qubits, error-correcting codes are typically employed to mitigate its effects. Interestingly, this redundancy not only helps with photon loss but also enhances the efficiency of Bell measurements~\cite{PhysRevLett.117.210501,Ewert_repeater_PRA,Schmidt_logical_BM,PhysRevA.104.052623,PRXQuantum.4.040322,patil2024improveddesignallphotonicquantum,PhysRevA.100.052303}. Additionally, the use of feed-forward techniques can further boost the efficiency~\cite{PhysRevA.100.052303}. This concept of `code-boosting' has already been applied in the FBQC proposal to improve its loss threshold~\cite{fbqc}, and recent studies have shown that combining code-boosting with feed-forward can drastically enhance the loss threshold~\cite{PhysRevLett.133.050605,PhysRevLett.133.050604}.

Recently, generalized fusion operations have been proposed that go beyond Bell-state measurements by projecting onto general, arbitrary maximally entangled states allowing for additional local rotations in these states. This leads to improved efficiencies when using one or two ancilla photons \cite{psiquantum_fusion}. It has also been shown that these generalized fusions, without ancilla states, are subject to the typical efficiency limit of \(\frac{1}{2}\), although they are no longer limited to standard Bell-state projections \cite{rimock2024generalizedtypeiifusion}. The generalized scheme proposed in Ref.~\cite{psiquantum_fusion}, which employs two single photons, achieves an efficiency of \(\frac{2}{3}\), whereas an ancilla-boosted Bell measurement with two unentangled ancilla photons only reaches an efficiency of 62.5\%~\cite{Ewert_BM,doi:10.1126/sciadv.adf4080}.
Inspired by these works, we aim to address the following questions: Are the schemes proposed in Ref.~\cite{psiquantum_fusion} already optimal? Furthermore, can these schemes be generalized to include more unentangled ancilla photons, and how does the success probability scale with the number of photons?

In addition, there are intriguing questions regarding the use of generalized fusions in the context of error-correcting codes. On the level of dual-rail qubits, it is possible to achieve better efficiencies by using generalized fusions instead of Bell measurements when ancilla photons are available. Is it also possible to find generalized fusions at the logical level of the error-correcting code that outperform logical Bell measurements? Moreover, how can these fusion schemes be realized?
It is well-known that boosted logical Bell measurements can be constructed from transversal boosted Bell measurements. Does this approach extend to the generalized fusion schemes proposed in Ref.~\cite{psiquantum_fusion}?
The aim of this work is to contribute towards answering some of the above-listed questions.

The paper is structured as follows.
In Sec.~\ref{sec:methods}, we discuss the methods used for our optimization tasks. Our results, first related with ancilla-boosted fusions and then concerning code-boosted fusions, are presented in Sec.~\ref{sec:results}, followed by the conclusion in Sec.~\ref{sec:conclusion}.

\section{Methods}\label{sec:methods}

To address the questions mentioned in the preceding section, we introduce a simple yet powerful toy model. This model generalizes known scenarios, such as ancilla-boosted physical Bell measurements~\cite{Grice_BM,Ewert_BM} and transversal ancilla-boosted Bell measurements within a logical Bell measurement of high-level error-correcting codes~\cite{Ewert_repeater_PRA,Schmidt_logical_BM}. Additionally, it extends to non-transversal logical Bell measurements, a domain that, to our knowledge, has not yet been explored.

\begin{figure}
    \centering
    \includegraphics[width=\linewidth]{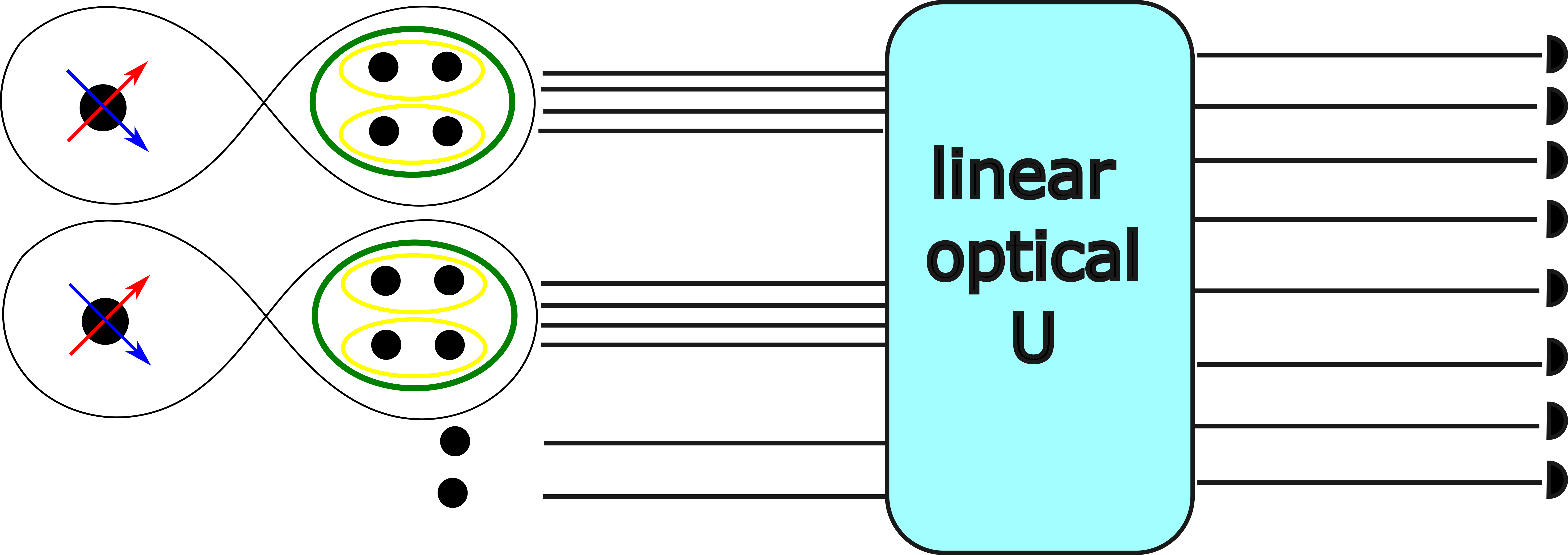}
    \caption{\justifying{The black dots with two colors represent generic qubits entangled with dual-rail qubits encoded using a QPC code (colored ellipses indicate the code stabilizers: yellow for $Z$-stabilizers and green for $X$-stabilizers). Additionally, an optional photonic ancilla state is positioned at the bottom. All optical modes are coupled through a linear-optical circuit $U$, and the outgoing modes are measured using photon-number-resolving detectors. We then select the events where the generic qubits are projected onto maximally entangled states.}}
    \label{fig:scheme}
\end{figure}

We start with the state
\begin{equation}
\frac{1}{2}\left(\ket{0}\ket{0^{(n,m)}}+\ket{1}\ket{1^{(n,m)}}\right)^{\otimes2} \otimes \ket{\text{ancilla}}\,,
\end{equation}
where $\ket{0}$ and $\ket{1}$ represent generic qubits, while $\ket{0^{(n,m)}}$ and $\ket{1^{(n,m)}}$ denote qubits encoded in the quantum parity code (QPC)~\cite{PhysRevLett.112.250501}, concatenated with the dual-rail qubit encoding~\cite{PhysRevLett.95.100501}.  In the dual-rail encoding, a qubit is encoded in the presence of a photon across two modes, 
\begin{align}
    \ket{0}=\ket{0,1}, \quad \ket{1}=\ket{1,0}\,.
\end{align}
The QPC encoding consists of two layers of repetition codes, thus generalizing the well-known Shor code with 3 blocks (2nd layer) and 3 qubits per block (1st layer)~\cite{PhysRevA.52.R2493}.  The first level protects against bit-flip errors, while the second one protects against phase-flips,
\begin{align}
    \ket{x}^{(m)}=\ket{x}^{\otimes m} \quad  x\in \mathbb{Z}_2\,,\\
    \ket{\pm}^{(n,m)}=\ket{\pm^{(m)}}^{\otimes n}\,,
\end{align}
where $\ket{\pm}=\frac{1}{\sqrt{2}}\left(\ket{0}\pm \ket{1}\right)$ applies on all encoding levels.
Thus, the overall QPC-encoded state (for one copy of the code) is a $2nm$-mode photonic state, where 
$\ket{\rm{ancilla}}$ is some additional photonic ancilla state.
In the next step, the modes of both QPC codes and the ancilla are coupled via static (passive) linear optics as it can be seen in Fig.~\ref{fig:scheme}. Formally, the vector of creation operators for the optical modes is transformed by a unitary matrix $U$. Finally, we perform photon counting on all optical modes and calculate the probabilities for all measurement outcomes, determining the corresponding post-measurement state of the two generic qubits.

For the remainder of the paper, we focus on the scenario where the post-measurement state of the generic qubits is maximally entangled. We do not consider any external loss, so all linear-optical operations are assumed to be ideal. All results presented were obtained using a Python simulator that we developed~\footnote{GitHub repository: \url{https://github.com/schmidtfrk/GeneralizedFusions}}, built on the QuTiP framework~\cite{qutip}.

To identify optimal linear optical transformations for effective fusion measurements, we employed the following optimization approach. Consider a QPC(\(n,m\))-encoding with \(j\) ancilla modes. In total, there are \(4nm + j\) modes on which we apply a linear optical operation, described by a unitary matrix \(U \in U(4nm + j)\), that transforms the mode operators. We consider the Clements decomposition~\cite{Clements} providing a surjective mapping
\begin{align}
   c: \mathbb{R}^{(4nm+j)(4nm+j-1)} \to U(4nm + j)\,, 
\end{align}
where the coefficients correspond to the phases and beam splitter transmission values in the Clements decomposition.

The overall optimization process consists of a cascade of individual optimizations. Since the probability of obtaining maximally entangled states is not continuous in \(U\), we introduce an auxiliary continuous objective function
\begin{align}
    f_x(U) = \sum_{y \in \Omega} p(y) E(\ket{\psi_y})^x    
\end{align}
to guide the optimization. Here, \(\Omega\) denotes the set of possible measurement outcomes, \(p(y)\) represents the probability of outcome \(y\) with the corresponding post-measurement state \(\ket{\psi_y}\), and \(E(\ket{\psi_y})\) is the entanglement entropy of the pure state \(\ket{\psi_y}\) of the two generic qubits.

The entanglement entropy of a state \(\ket{\psi}_{AB}\) is defined as:
\begin{align}
        E(\ket{\psi}_{AB}) = -\Tr\left(\rho_A \log\left(\rho_A\right)\right)\,,
\end{align}
where
\begin{align}
    \rho_A = \Tr_B \left(\ket{\psi}_{AB}\bra{\psi}_{AB}\right)\,.    
\end{align}
The parameter \(x\) is a free variable that controls the weighting of entanglement in the optimization. As \(x\) increases, the optimization of \(f_x(U)\) becomes more biased toward strongly entangled states. However, for a random unitary \(U\) and large \(x\), \(f_x(U)\approx0\) with little sensitivity to small variations in \(U\) with high probability. By contrast, smaller values of \(x\) result in larger variations. Therefore, we perform multiple rounds of optimization, gradually increasing the value of \(x\).

We begin with an initial unitary \(U_0\) and, to avoid convergence to a local optimum, generate several perturbed versions of \(U_0\). We first identify a \(\tilde{U}_0 \in \mathbb{R}^{(4nm+j)(4nm+j-1)}\) such that \(c(\tilde{U}_0) = U_0\). Then, we create a set of perturbed unitaries:
\begin{align}
    \{U_0^{(k)} = c\left(\tilde{U}_0 + \mathcal{N}_k(0, \sigma^2)\right)\}\,,    
\end{align}
where \(\mathcal{N}_k(0, \sigma^2)\) is a vector of independent Gaussian random variables with mean 0 and variance \(\sigma^2\). We then perform numerical optimization of \(f_x\) using the perturbed unitaries \(U^{(k)}_0\) as starting points.

The unitary that yields the best optimization result is selected as \(U_1\), which serves as the starting point for the next round of optimization. This process is repeated for multiple rounds until \(f_x\) with \(x \gg 1\) converges with respect to the number of optimization rounds. Typically, convergence is achieved after 3-4 rounds.

We are also interested in whether global Bell measurements can outperform transversal ones when applied to a stabilizer code. To explore this, we introduce a target function \(B_x(U)\), which we refer to as `Bellness':
\begin{align}
    B_x(U) = \sum_{y \in \Omega} p(y) \sum_{k, l = 0}^1 |\bra{\Phi_{k,l}} \ket{\psi_y}|^x \,,    
\end{align}
where \(\ket{\Phi_{k,l}} = \frac{1}{\sqrt{2}} \left( \ket{0,k} + (-1)^l \ket{1,1-k} \right)\) are Bell states. In the limit as \(x \to \infty\), this target function aims to maximize the probability of projecting the conditional generic qubit states onto Bell states~\footnote{Equivalently, the linear optical measurement performs a projective Bell measurement on the two measured qubits.}. 
Similar to the previous approach, we consider a sequence of optimizations to enhance convergence.

\section{Results}\label{sec:results}

In this section, we will first investigate ancilla-boosting and subsequently code-boosting with the QPC encoding,
presenting the results that we obtained based on our numerical optimization.

\subsection{Ancilla-boosting}
\begin{figure}
    \centering
    \includegraphics[width=0.75\linewidth]{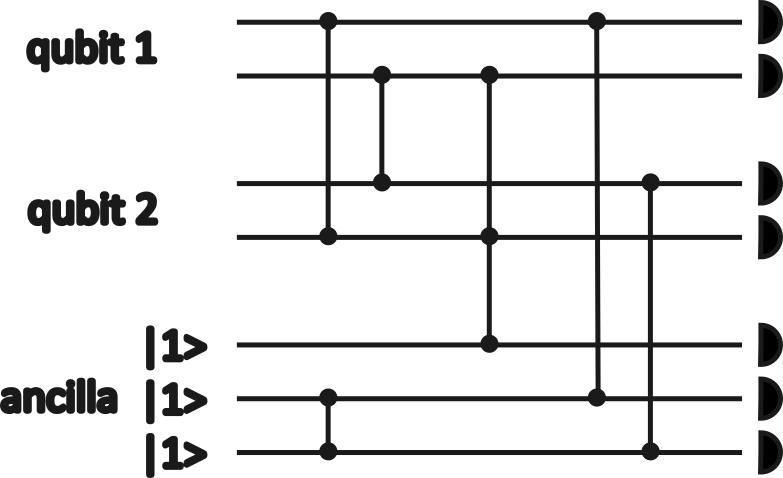}
    \caption{\justifying{Linear-optical seven-mode circuit (as $4nm=4$ and $N=3$) of our proposed generalized fusion scheme utilizing three single ancilla photons. The interactions between the modes are described by the matrix of a discrete Fourier transfrom (DFT) and the large dots denote the modes involved in each interaction. Note that a two-mode DFT operation is a simple 50:50 beam splitter. }}
    \label{fig:3_photon_scheme}
\end{figure}

As a first step, we address the earlier questions regarding fusion operations without any additional QPC encoding \(((n,m) = (1,1))\), focusing on different ancilla states. The relevant simulation code is available at GitHub~\footnote{GitHub repository: \url{https://github.com/schmidtfrk/GeneralizedFusions}}.

The simplest scenario is the case without any additional ancilla modes. It is well-established that fusion operations based on the unambiguous discrimination of Bell states cannot achieve an efficiency greater than \(\frac{1}{2}\)~\cite{Calsamiglia2001}. Our numerical simulations quickly converged to \(\frac{1}{2}\), strongly suggesting that more general fusion measurements do not yield a higher success probability. This conclusion has been independently confirmed in a recent study~\cite{rimock2024generalizedtypeiifusion}, 
where the authors provided an analytical proof under the assumptions of static linear optics, no ancilla photons, and no other redundancy such as extra photons in an error-correcting code.
\begin{figure}
    \centering
    \begin{subfigure}{0.45\linewidth}
         \includegraphics[width=\linewidth]{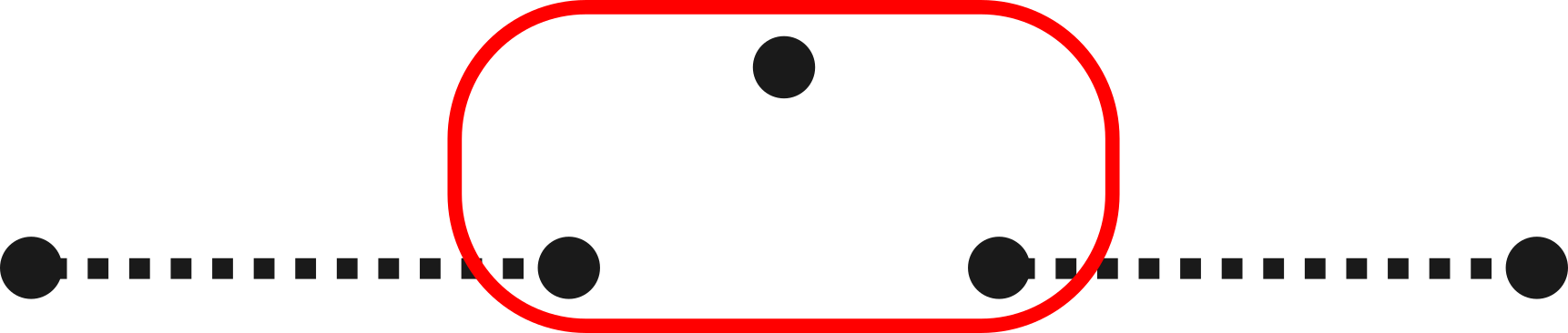}
         \caption{}
    \label{fig:comparison_1ancilla}   
    \end{subfigure}
        \begin{subfigure}{0.45\linewidth}
         \includegraphics[width=\linewidth]{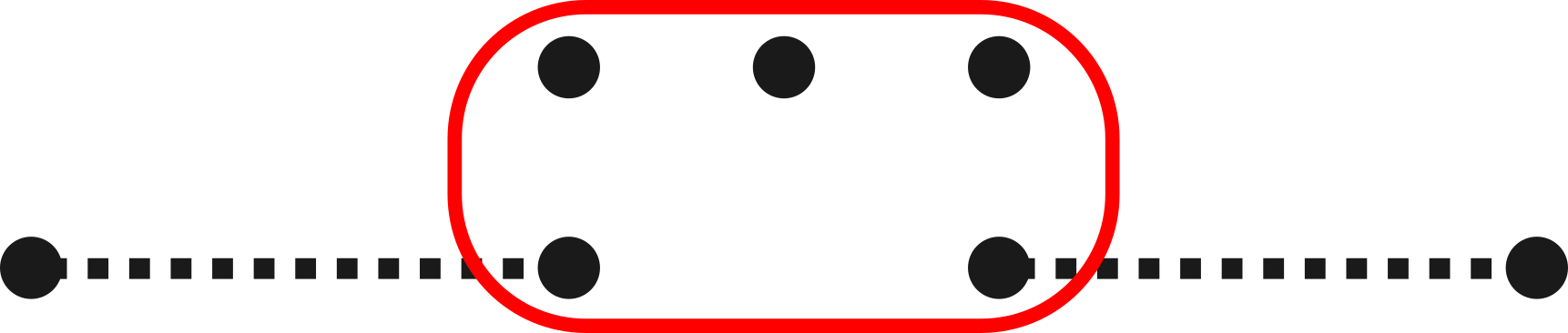}
         \caption{}
    \label{fig:comparison_3ancilla}   
    \end{subfigure}\\
    
    \begin{subfigure}{0.45\linewidth}
         \includegraphics[width=\linewidth]{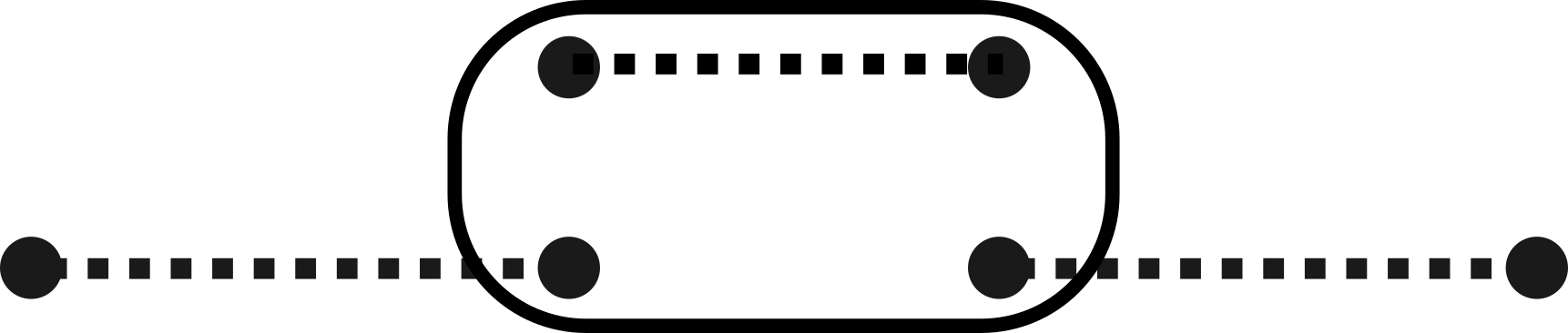}
         \caption{}
    \label{fig:comparison_grice}   
    \end{subfigure}
    \begin{subfigure}{0.45\linewidth}
         \includegraphics[width=\linewidth]{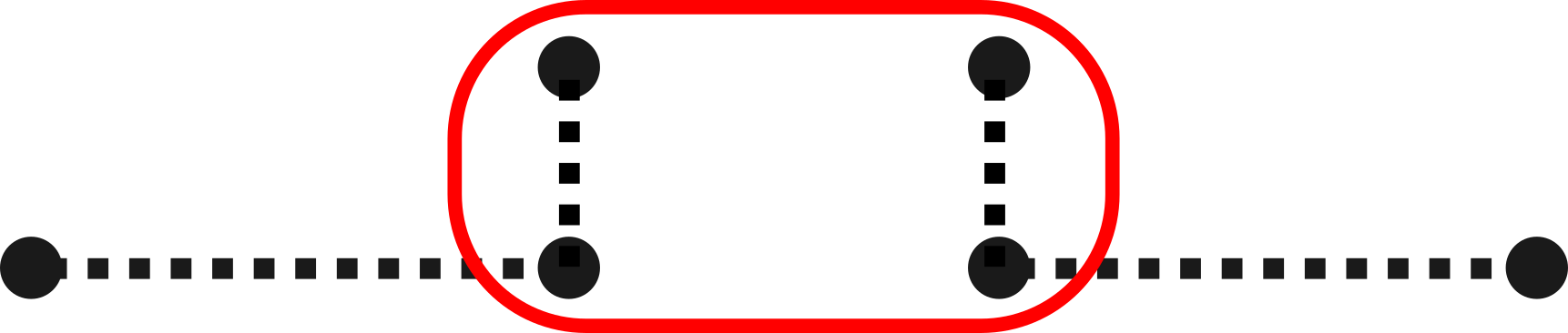}
         \caption{}
    \label{fig:comparison_qpc21_generalized}   
    \end{subfigure}\\
        \begin{subfigure}{0.45\linewidth}
         \includegraphics[width=\linewidth]{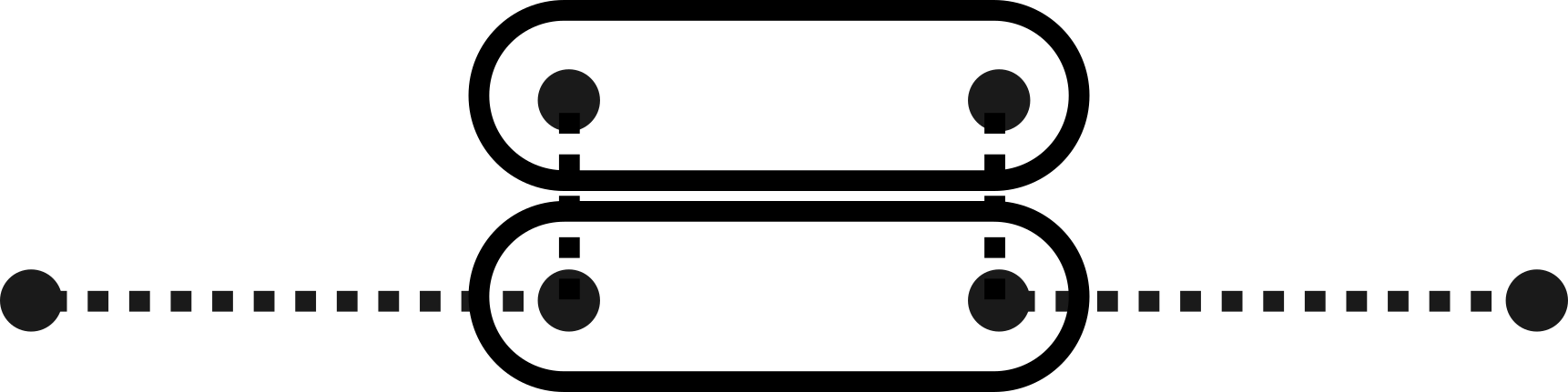}
         \caption{}
    \label{fig:comparison_qpc21_trans_bm}   
    \end{subfigure}
           \begin{subfigure}{0.45\linewidth}
         \includegraphics[width=\linewidth]{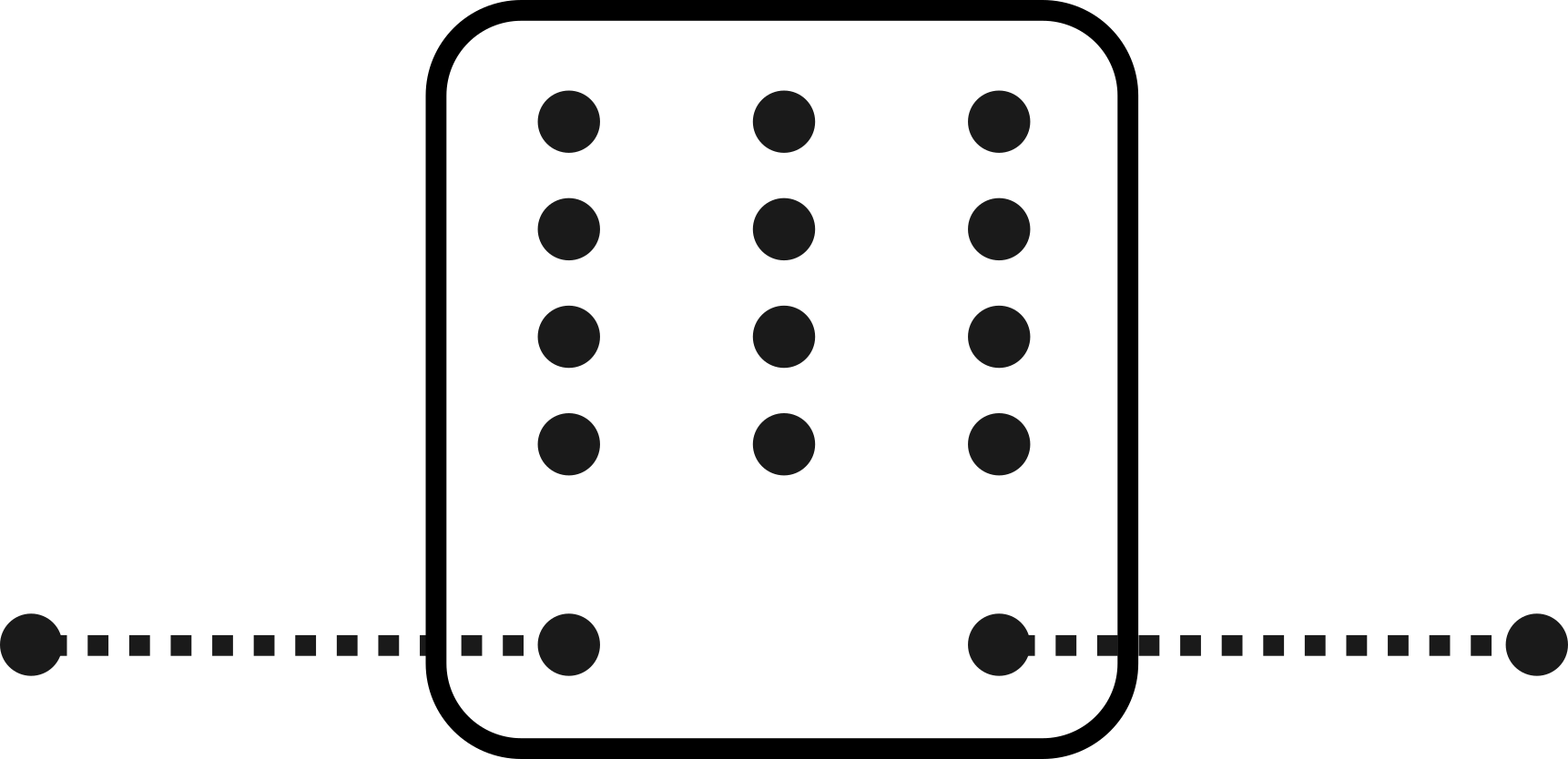}
         \caption{}
    \label{fig:comparison_ewert_12_ancilla}   
    \end{subfigure}
           \begin{subfigure}{0.45\linewidth}
         \includegraphics[width=\linewidth]{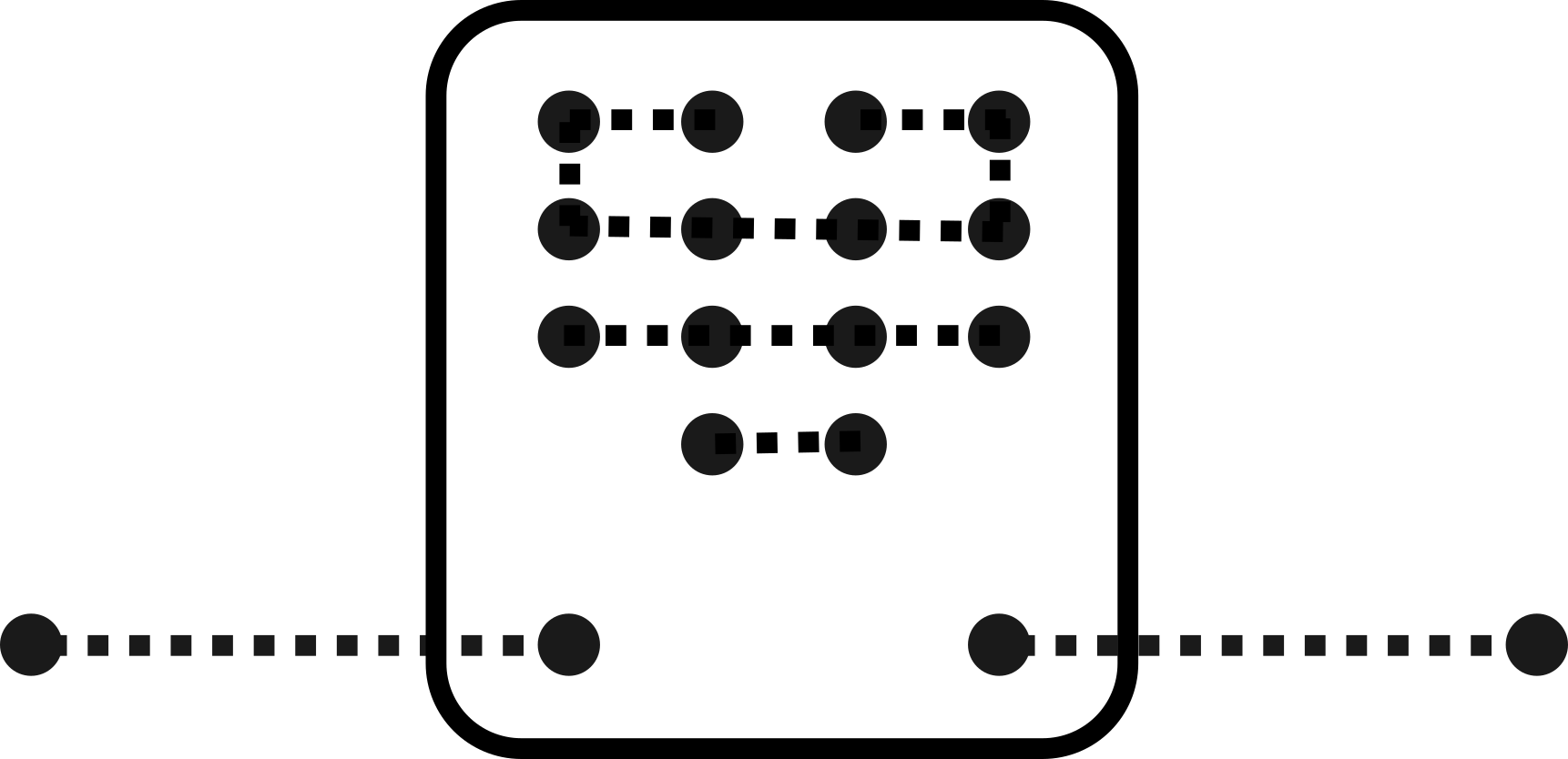}
         \caption{}
    \label{fig:comparison_grice_14_ancilla}   
    \end{subfigure}
    \caption{\justifying{Illustration for comparing resource costs (number of redundant or redundantly encoded photons, offline entangled photons) and types of fusion (red: generalized fusion, black: Bell-type measurement). The dashed lines denote entanglement between different systems. (a) ancilla-boosted generalized  fusion with an efficiency of 7/12~\cite{psiquantum_fusion}(b) ancilla-boosted generalized  fusion with an efficiency of 17/24 (c) ancilla-boosted Bell measurement with an efficiency of 3/4~\cite{Grice_BM}
    (d) code-boosted generalized fusion with an efficiency of 13/16 (e) code-boosted Bell measurement with an efficiency of 3/4~\cite{PhysRevLett.117.210501}
    (f) ancilla-boosted Bell measurement with an efficiency of 25/32~\cite{Ewert_BM} (g) ancilla-boosted Bell measurement with an efficiency of 15/16~\cite{Grice_BM}}.}
    \label{fig:ressources}
\end{figure}

We now explore a slightly different scenario by allowing ancilla states composed of multiple single photons (i.e., $\ket{\text{ancilla}}=\ket{1}^{\otimes N}$, where $N$ denotes the number of single photons) and examining whether the schemes employing one and two auxiliary single photons, as proposed in Ref.~\cite{psiquantum_fusion}, with reported efficiencies of \(7/12\) and \(8/12\), are indeed optimal. Our numerical simulations suggest that these schemes are indeed optimal, as our optimization consistently converged to similar configurations. While the linear-optics unitary matched for the ancilla modes, it differed in other modes, indicating that some degrees of freedom remain in the circuit without affecting the optimal efficiency.
Moreover, our simulations were unable to achieve the efficiencies of \(7/12\) and \(8/12\) using boosted Bell measurements instead of generalized fusions.

As the next step, we considered the case with three single auxiliary photons. Our optimization process yielded an efficiency of \(17/24\). Despite performing multiple optimization runs, each converging to the same efficiency, the resulting circuits varied, even after accounting for negligible phase differences. 
Nonetheless, a possible decomposition of this scheme into 50:50 beam splitters and a three-mode tritter is shown in Fig.~\ref{fig:3_photon_scheme}.
Interestingly, with four extra single photons, we achieved an optimized efficiency of \(3/4\). Notably, this efficiency matches that obtained by the boosted Bell measurement scheme proposed in Ref.~\cite{Ewert_BM}.

Additionally, we observed an intriguing pattern in the efficiencies of generalized fusion measurements. Starting with no ancilla photons, the efficiency is \(1/2\). Introducing a single ancilla photon raises the efficiency to \(7/12 = 1/2 + 1/12\). Adding a second photon further increases the efficiency to \(2/3 = 1/2 + 2/12\). With a third photon, the efficiency improves to \(17/24 = 2/3 + 1/24\). Finally, with four photons, we achieve an efficiency of \(3/4 = 2/3 + 2/24\).

Based on this pattern, we conjecture that with five extra single photons, the efficiency will be \(37/48 = 3/4 + 1/48\). However, for five single photons, our simulations do not consistently converge to a single efficiency value when starting from random initialization. It appears that the optimization often converges to configurations that effectively utilize fewer than five photons, leaving the remaining ones unused.

To address this, we are currently working on understanding the smaller schemes in greater detail, allowing us to start from more informed initial configurations rather than random ones. Furthermore, we conjecture that for an even number of ancilla photons \(2z\), the success probability of the generalized fusion is given by
\begin{align}\label{eq:conjecture_even}
\frac{1}{2} + \frac{1}{3}\sum_{j=1}^z 2^{-j} = \frac{5}{6} - \frac{2^{-z}}{3}\,.    
\end{align}
For an odd number of photons \(2z+1\), the probability is given by
\begin{align}\label{eq:conjecture_odd}
\frac{1}{2} +\frac{1}{3} \sum_{j=1}^z 2^{-j} + \frac{2^{-z}}{12} = \frac{5}{6} - \frac{2^{-z}}{4}\,.    
\end{align}
There are several schemes for boosted Bell measurements that achieve efficiencies approaching unity by utilizing entangled ancilla states~\cite{Grice_BM,Ewert_BM}. Conceptually, it is intriguing that the use of entangled ancilla states appears to be not just advantageous but necessary, as achieving unit efficiency seems impossible without them.

Reference~\cite{Ewert_BM} also explored whether the boosted Bell-measurement efficiency of \(3/4\) (corresponding to \(z=2\) in Eq.~(\ref{eq:conjecture_even})) can be surpassed by incorporating additional unentangled photons. The authors proposed a scheme utilizing 12 single photons, achieving an efficiency of \(25/32 = 78.125\%\)~\cite{Ewert_BM}, which falls short of our conjectured upper bound of \(53/64 \approx 82.8\%\).
Our conjectured upper bound also serves as a constraint on the efficiencies of Bell measurements, which were analyzed in Ref.~\cite{PhysRevA.98.042323}, assuming dual-rail qubits encoded in the polarization of photons. Interestingly, their analytical upper bounds closely align with our conjecture. Due to the structure of their proof, these bounds assume polarization-preserving interferometers. In the scenario where the initial polarization of some or all modes may be rotated by $\pi/4$, the authors of Ref.~\cite{PhysRevA.98.042323} derived an upper bound of $54/64$, which, according to our conjecture, may not be tight. Additionally, for networks where the two Bell states are first interfered on a 50:50 beam splitter and each half analyzed separately, they reported an upper bound of $52/64$.
Thus, to close the gap between this bound and our bound, also depending on the application, either a more general Bell-measurement circuit (which may not even exist for attaining our bound) or a generalized fusion is needed.  ´
A little illustrative summary and comparison of different linear optical fusions and Bell measurements with respect to their used ressources is given in Fig.~\ref{fig:ressources}.

\subsection{Code-boosting}

Let us now address the questions related to the additional QPC encoding. It is well-known that logical Bell measurements on stabilizer codes can be decomposed into Bell measurements on each qubit pair. This decomposition has been widely utilized~\cite{PhysRevLett.117.210501,Ewert_repeater_PRA, Schmidt_logical_BM, PhysRevA.100.052303} in the context of linear optical Bell measurements, where the specific choice of individual Bell measurements can lead to varying efficiencies~\cite{Schmidt_logical_BM}.

Typically, the efficiency of a logical Bell measurement is superior to that of a single dual-rail qubit Bell measurement in the absence of photon loss, though there are exceptions (see~\cite[p.8]{Schmidt_logical_BM}). We explored applying generalized fusions with a single ancilla photon, as proposed in~\cite{psiquantum_fusion}, in a transversal manner to a QPC(2,1) encoding. We found that this scheme achieves an efficiency of approximately \(67.4\%\), which is lower than the efficiency of a Bell measurement on a single, redundantly encoded QPC(2,1) qubit pair without any additional ancilla photons. This suggests that generalized fusions may not be particularly beneficial when applied transversally.
Using simple Bell measurements with the QPC(2,1) encoding can yield an efficiency of \(3/4\) without any additional external ancilla photons. This efficiency can be achieved using static linear optics, where for the standard scheme~\cite{PhysRevLett.117.210501}, the efficiency is given by \(1 - 2^{-n}\), or by \(1 - 2^{-(n+m-1)}\) in a more advanced scheme~\cite{Schmidt_logical_BM}\footnote{As this scheme always provides information about the \(YY\) observable, the enhanced efficiency may not always be useful, as many applications such as FBQC would typically require information about the \(X\)- or \(Z\)-eigenstates.}. Alternatively, with the use of feed-forward, the efficiency is further improved to \(1 - 2^{-nm}\)~\cite{PhysRevA.100.052303}, which, however, does not provide a higher efficiency beyond the static scheme for QPC(2,1) or generally QPC($n$,1).

We also explored a general global fusion measurement for QPC(2,1) without any extra ancilla photons and found that our optimization converges to \(13/16\). This result provides the first example of a stabilizer encoding where (global) generalized fusions outperform (transversal) Bell measurements. While it is unclear whether global Bell measurements can surpass transversal Bell measurements, generalized fusions can serve as an upper bound. As QPC(2,1) allows to detect phase-flip errors and as its extensions to QPC($n$,1) repetition codes also allow to correct such errors, the use of these codes for fusion-based schemes can be beneficial, and so code-boosting the fusion efficiencies can have the additional benefit of error correction \cite{fbqc,PhysRevLett.117.210501}. While our example is only for QPC(2,1), improved, generalized, global, code-boosted fusions may exist for larger codes as well such as QPC($n$,1), certainly QPC(1,2) and QPC(1,$m$), and possibly even QPC($n$,$m$) which is useful for loss correction.    
 
\section{Conclusion}\label{sec:conclusion}

We conducted numerical simulations and optimizations to investigate the efficiencies of generalized fusion measurements using static linear optics. These generalizations encompass ancilla-boosting, code-boosting, and the exploration of measurements beyond Bell measurements. For ancilla-boosting alone, i.e., for standard dual-rail qubits without any extra redundantly encoded photons, when employing multiple single photons as the ancilla state and extending the measurements beyond Bell measurements, we identified a pattern in the efficiencies for up to four photons. We conjecture that this pattern persists, apparently implying that entangled ancilla states are necessary to achieve near-deterministic measurements. 

This raises several open questions: Can our conjecture be formally proven? There are at least two (similar) known schemes that utilize entangled ancilla states to achieve near-deterministic Bell measurements, but how much entanglement is truly required? Notably, only with four single photons was it possible to attain the highest generalized fusion efficiency using a Bell measurement. In which scenarios do these measurements yield the same optimal efficiencies? In which cases are generalized fusions that project onto maximally entangled states, rather than Bell states, advantageous? In the context of simple dual-rail encoded qubits, Bell states and maximally entangled states only differ by linear optical transformations, while in the case of an additional error-correcting code, which is needed for code-boosting, the differences are not necessarily easy to remove. Moreover, many applications of entangling measurements make use of the stabilizer formalism that is directly compatible with Bell measurements, but within this common formal framework it is not straightforward to use generalized fusions instead.

In the specific context of code-boosting beyond Bell measurements, we explored the QPC(2,1) encoding as a simple example. We discovered a generalized fusion scheme that surpasses the best-known efficiency achieved with transversal Bell measurements, even when incorporating feed-forward, as for such instances of the QPC encoding (with, more specifically, only one physical qubit per code block or all physical qubits inside a single code block), the obtainable efficiencies of the static and the dynamic Bell measurement schemes coincide. However, we also examined the potential benefits of transversal generalized fusion measurements and found that they do not generally offer an advantage. 

As for an outlook, it is of interest for applications in quantum computation (MBQC/FBQC) and quantum communication (such as quantum repeaters employing quantum error correction codes) whether improved, code-boosted entangling and fusion measurements can be beneficial for both enhancing fusion efficiencies and correcting errors -- beyond what is already known for Bell measurements. In particular, photon loss is known to be correctable together with code-boosting for more general QPC encoding and other codes using Bell measurements. While here we have focused on generalized fusions in an ideal error-free scenario, providing some contributions with nonetheless still several remaining open questions even in this constrained setting, an obvious next and important step would be to incorporate errors.   

\section{Acknowledgments}
We thank the BMBF in Germany for funding via the project PhotonQ.

 \bibliography{apssamp}

\end{document}